# Directional ballistic transport in the two-dimensional metal PdCoO$_2$

Maja D. Bachmann[1,2,†,*], Aaron L. Sharpe[3,4,†,*], Arthur W. Barnard[5], Carsten Putzke[1,6], Thomas Scaffidi[7], Nabhanila Nandi[1], Seunghyun Khim[1], Markus König[1], David Goldhaber-Gordon[4,5], Andrew P. Mackenzie[1,2,*], Philip J. W. Moll[1,6,*]

[1]Max Planck Institute for Chemical Physics of Solids, 01187 Dresden, Germany
[2]School of Physics and Astronomy, University of St Andrews, St Andrews KY16 9SS, UK
[3]Department of Applied Physics, Stanford University, 94305 Stanford, California, USA
[4]SLAC National Accelerator Laboratory, 94025 Menlo Park, California, USA
[5]Department of Physics, Stanford University, 94305 Stanford, California, USA
[6]Institute of Materials, École Polytechnique Fédérale de Lausanne (EPFL), 1015 Lausanne, Switzerland
[7]Department of Physics, University of Toronto, Toronto, Ontario M5S 1A7, Canada

[†]These authors contributed equally to this work.
*correspondence address: maja.bachmann@stanford.edu, aaron.sharpe@stanford.edu, andy.mackenzie@cpfs.mpg.de, philip.moll@epfl.ch

**In an idealized infinite crystal, the material properties are constrained by the symmetries of its unit cell. Naturally, the point-group symmetry is broken by the sample shape of any finite crystal, yet this is commonly unobservable in macroscopic metals. To sense the shape-induced symmetry lowering in such metals, long-lived bulk states originating from anisotropic Fermi surfaces are needed. Here we show how strongly facetted Fermi surfaces and long quasiparticle mean free paths present in microstructures of PdCoO$_2$ yield an in-plane resistivity anisotropy that is forbidden by symmetry on an infinite hexagonal lattice. Bar shaped transport devices narrower than the mean free path are carved from single crystals using focused ion beam (FIB) milling, such that the ballistic charge carriers at low temperatures frequently collide with both sidewalls defining a channel. Two symmetry-forbidden transport signatures appear: the in-plane resistivity anisotropy exceeds a factor of 2, and transverse voltages appear in zero magnetic field. We robustly identify the channel direction as the source of symmetry breaking via ballistic Monte-Carlo simulations and numerical solution of the Boltzmann equation.**

INTRODUCTION

The directionality, or anisotropy, of the electrical resistivity of a crystalline material is determined by the point group symmetry of its underlying lattice. The resistivity $\rho$ in two dimensions for square, triangular and hexagonal lattices is isotropic (see Supplementary Note 1). Only if the rotational symmetry is lowered to two-fold is in-plane anisotropy permitted, such that the two diagonal components $\rho_{xx}$ and $\rho_{yy}$ differ. Such symmetry-lowering has attracted significant attention due to the study of so-called electronic nematic and smectic liquid crystals, in which self-organization of the electron fluid is thought to be the driver of the broken symmetry[1-5]. Indeed, transport measurements sensitive to resistive anisotropy have been one of the key probes of this class of physics. In this work we address the question of whether other approaches can also induce transport anisotropies. Specifically, we investigate whether there are observable consequences of effective symmetry breaking on the micro-scale by creating devices of differing orientation relative to an underlying crystal lattice whose point group symmetries are not altered by the fabrication process.

Fundamentally, all bulk crystalline symmetries are broken in any metallic microstructure due to its finite size. However, this has no observable effects in the bulk transport in most cases for the simple reason that the material is in the usual 'ohmic' or 'diffusive' transport regime in which the carriers scatter so strongly in the bulk that the boundaries are irrelevant. In this case, the current and electric field distribution is well described by a resistivity tensor $\rho$ that adheres to the crystalline point group symmetry. It has been known for decades that it is possible to purify metallic crystals enough to enter the so-called 'ballistic' transport regime, in which the electron mean free path $\lambda$ between internal scattering events exceeds the minimum sample dimension. In this regime, boundary scattering becomes relevant or even dominant. However, this alone is not sufficient to produce observable resistivity anisotropy. The essential additional ingredient is significant Fermi surface anisotropy. Early theoretical consideration of ellipsoidal Fermi surfaces[6-9] led to predictions of transport anisotropies, but experiments in aluminium[10,11] did not resolve such effects. Recent results on epitaxial tungsten thin films have detected a growth-direction dependence of the resistance when the films are thin enough to be in the ballistic limit in the direction perpendicular to the substrate[12]. This result is attributed to the anisotropy of the three-dimensional Fermi surface of tungsten, and hence suggests that boundary-induced symmetry breaking is achievable. Here, we exploit the in-plane anisotropy of the Fermi surface in a two-dimensional metal, $PdCoO_2$, to demonstrate not only that directional symmetry breaking is achievable, but that it can be a large effect. By cutting our directional channels from the same single crystal we remove any sample-dependent uncertainties from the experiments. We present a simple intuitive picture to explain our observations, and then reinforce it with calculations and Monte Carlo simulations.

EXPERIMENT

The ultra-clean, naturally layered crystal structure of $PdCoO_2$ is host to extremely conductive, quasi two-dimensional sheets of palladium, separated by layers of $CoO_2$ octahedra. Due to the strikingly high purity of this oxide[13], it can support electron mean free paths of up to 20 μm at temperatures below 20K, as evidenced by the in-plane residual resistivity value of only 8 nΩcm[14]. Extensive de Haas-van Alphen (dHvA)[15], angle resolved photoemission (ARPES)[16,17] and angle-dependent magnetoresistance (AMRO)[18] measurements have well characterized its hexagonal Fermi surface, which fills half of the Brillouin zone, as expected for a monovalent metal. The out-of-plane dispersion is so weak that transport is essentially two-dimensional[19] and we therefore work here in a 2D approximation. The extremely long in-plane mean free path of $PdCoO_2$ has been directly demonstrated in measurements of transverse electron focusing[20] and the observation of field-periodic oscillations in microstructures[21].

The crystals of $PdCoO_2$ used in this study grow as thin platelets, with a typical thickness of approximately 5-30 μm and lateral dimensions of several hundred micrometers. Despite the layered character of the crystal structure, the crystals do not exfoliate, which obstructs the use of commonly used clean room nanofabrication techniques. Instead, we employ a focused ion beam (FIB) for three-dimensional micro-sculpting of as-grown crystals. Details of crystal synthesis[20] and FIB microstructuring[20,22] are given elsewhere.

A typical $PdCoO_2$ transport device produced by FIB micromachining is displayed in Fig. 1a. Conveniently, the growth edges of the $PdCoO_2$ crystals are oriented perpendicular to the crystallographic axes, so that the crystal orientation can easily be determined. This permits the fabrication of four serial transport bars within the same single crystal, which are precisely oriented with respect to the crystal lattice. Due to the in-plane 6-fold rotational symmetry of the palladium planes, the full angular range can be spanned in steps of 10°, by choosing to measure parallel to the crystal axis (0°) as well as 10°, 20° and 30° off axis.

RESULTS

Directional ballistic effects

The angular dependence of the in-plane mean free path $\lambda$ of the oriented bars of $PdCoO_2$ microstructure of Fig. 1a is presented as a function of temperature in Fig. 1b, in comparison with that of a bulk (155 μm wide) channel of the same orientation. Data are also shown for the same bar after subsequent narrowing from 7μm to 2.5μm width, and are seen to evolve in 'text-book' fashion: the temperature at which the data first deviate from the diffusive regime is width-dependent (see Supplementary Note 2 for details) and the low temperature value of $\lambda$ for the restricted channels is limited by their widths rather than by bulk scattering. At high temperatures, $\lambda$ is strongly limited by phonon-scattering and the sample is in a diffusive transport regime, hence its transport is isotropic. In contrast, at low temperatures, once the electron mean free path exceeds the width of the transport bars, a pronounced anisotropy appears. In particular, for the device displayed in Fig. 1a in which the bars are 7μm wide, the crystal-symmetry-forbidden resistivity anisotropy $(\rho_{30} - \rho_0)/\rho_0$ is as large as 50% between the highest and lowest resistive direction. Upon thinning down the bars to 2.5μm width, the ratio further increased to 200%.

The order of the curves, from least to most conductive, can be understood qualitatively by considering Fig. 2. When the transport bar is oriented such that it is aligned with one of 3 main directions of the Fermi velocity a large number of electronic states propagate parallel to the bar and avoid any surface collisions. On the other hand, when the orientation of the transport bar is rotated by 30°, the dominant ballistic directions guide the electrons towards the sample edges, leading to frequent boundary scattering events.

These results clearly demonstrate the notion of *directional ballistics*. In any material with a circular Fermi surface the apparent resistivity is enhanced due to boundary scattering, depending on the specularity of the boundaries[23]. This effect is isotropic: a bar of a given width and length will have the same resistance no matter what orientation it is cut in. Most high-mobility two-dimensional electron gases have circular or smoothly evolving Fermi surfaces, in which no orientation dependence is observable. In contrast, a Fermi surface with a strongly non-isotropic Fermi velocity distribution can significantly modify the rate of boundary scattering and therefore support an orientation dependence of the resistance. A comparison of the Fermi surface shape and the velocity density map highlights the subtle role of anisotropy (Fig. 3). The overall 2D Fermi surface of $PdCoO_2$ does not deviate much from a circular approximation, and given its 6-fold rotational symmetry, it may not strike the eye as particularly anisotropic. Indeed, the magnitude of the Fermi velocity of $PdCoO_2$ is almost constant around the Fermi surface[14]. The key aspect of directional ballistics, however, is the strong angle dependence of the velocity direction distribution, i.e. the probability of finding a certain direction of quasiparticle velocity (see Supplementary Note 6). As the nearly-flat Fermi surface segments host large densities of states propagating essentially into the same direction, the velocity direction distribution is extremely anisotropic despite the relatively isotropic appearance of the Fermi surface. We propose such a velocity direction distribution map as a tool to visualize the propensity of a material to exhibit directional ballistics.

Transverse voltages in zero field

Thus far we have been concerned with directional ballistic effects observable in the longitudinal electrical transport. However, due to the broken rotational symmetry at the boundaries, finite off-diagonal terms are allowed in the conductivity matrix along low-symmetry directions. Such terms have been used to effectively probe bulk symmetry-lowering in $Ba(Fe_{1-x}Co_x)_2As_2$ single crystals[24] and $La_{2-x}Sr_xCuO_4$ thin films[25]. In $PdCoO_2$, these are expected due to directional microstructuring and can be accessed by the device geometry outlined in Fig. 4a. Along a low symmetry direction, a transverse voltage in the ballistic regime develops due to an imbalance of electrons propagating towards the two different sides of the transport channel. Consequently, one expects the transverse voltage to be of equal strength but opposite sign with respect to the angle tilted away from a high symmetry direction.

This can be tested in a specifically designed microstructure of $PdCoO_2$ (Fig. 4b). The outline of the crystal is indicated by a dashed line. Following long current homogenizing meanders at the injection, the heart of the sample consists of two serial transport bars cut at +3° and -3° with respect to the 0°degree direction, both 2.2μm wide. Each of the bars has 3 pairs of opposite voltage contacts, allowing for simultaneous transverse and longitudinal voltage measurements. The resulting temperature dependence of the transverse zero-field resistivities $\rho_{xy}$ is presented in (Fig. 4c). In the diffusive transport regime at high temperatures, the transverse resistivity is absent, as expected. With the dominant scattering off phonons in the bulk, the point-group symmetry of the material dominates the scattering and hence in-plane isotropy is symmetry enforced. Yet upon entering the ballistic regime, a finite, asymmetric voltage develops across the device depending on the orientation of the transport bar.

ANALYSIS

In order to estimate the expected magnitudes of the effects described above, we have taken two theoretical approaches, each with their own advantages and drawbacks. Firstly, we use ballistic Monte Carlo simulations based on Landauer-Büttiker formalism (see Supplementary Note 4). In our implementation of this technique, we can incorporate a realistically parametrized Fermi surface such as that presented in Fig. 2a. However, our code does not take account of bulk impurity scattering which, while small, still has relevance in the real devices as they are much longer than the mean free path. For this reason, we also analyzed standard device geometries using numerical kinetic calculations within the Boltzmann equation formalism (see Supplementary Note 5). In this case it was possible to include bulk scattering, but only using an idealized perfectly hexagonal Fermi surface. Neither calculation is therefore expected to be a perfect numerical match to the data, and both can be expected to overestimate the anisotropy and the size of the transverse voltage. This is indeed the case: The Monte Carlo simulation predicts longitudinal anisotropy factors of 4.5 and 7, and the kinetic calculations 1.9 and 5 for 7 µm and 2.5 µm wide channels respectively, compared with experimentally measured values of 1.6 and 2.4. We note that the Monte Carlo simulation predicts the width dependence of the anisotropy well, showing that it rises by a factor of 1.6 between widths of 7 µm and 2.5 µm. This is significant because, in line with intuition, the simulation further predicts continued growth of this anisotropy as the width is decreased.

In spite of the above caveats, the overall agreement between directional ballistic theory and experiment is good, and highlights the importance of taking the Fermi surface shape and channel direction into account in the analysis of data from width-restricted channels of materials such as $PdCoO_2$ with faceted Fermi surfaces. The analysis presented in reference[26] considered a circular Fermi surface and hence the orientation of the channel relative to the crystal axes to be unimportant. The results presented here show that to conclusively identify a viscous contribution to transport in $PdCoO_2$ further experiments on transport bars aligned both along the 0° and 30° directions will be required, combined with analysis using models in which the faceting of the Fermi surface is taken into account[27-29].

CONCLUSIONS AND OUTLOOK

We have shown, for the first time in a two-dimensional metal, that a strongly faceted Fermi surface can lead to strongly orientation dependent conduction in otherwise identical ballistic devices cut from the same single crystal. These observations are of fundamental and practical importance to the question of the minimal attainable resistance in nanoscopic conductors, which ultimately limits the potential miniaturization of electric conductors in technological applications. As conductors are scaled down, even technologically relevant thin films enter the ballistic transport regime at elevated temperature, and boundaries become an important source of scattering[12]. Our results demonstzrate that the boundary scattering contribution in zero field can be reduced by over a factor of two when a 2.5 µm wide channel is aligned with one of the main directions of quasiparticle propagation. This is not a fundamental limit; indeed the minimum width we have studied in these proof-of-principle experiments is at least an order of magnitude larger than the minimum that could be envisaged. For wires less than 10 µm long, narrowed to widths of order 100 nm, the effect may be much larger, resulting in significant gains in attainable channel conductivity compared to that available from materials with circular Fermi surfaces. Our results invite investigation of other delafossite metals in which there are subtle differences in the degree of Fermi surface faceting[19], and also a thorough study of the effects of magnetic field on directional ballistics.

Finally, we note that the phenomena we report here are far from being restricted to delafossites. Materials with facetted Fermi surfaces are not rare, with gated bilayer graphene[30] being one of the most promising platforms for extensions of this research. Most strikingly, the in-plane transport anisotropy is a widely used technique to detect subtle symmetry-lowering electronic states such as electronic nematicity. The implicit assumption is based on the group-theoretical argument that an in-plane anisotropy necessitates a rotational symmetry being lowered to two-fold. Our work shows that especially in clean metallic crystals even the symmetry lowering by the sample shape itself can induce such an anisotropic response, in the absence of a two-fold symmetry. As mean free paths can indeed become macroscopic, these effects may well appear in traditional single crystals. This may prove to be important in the interpretation of unconventional transport phenomena in topological semi-metals, which generally tend to be of high mobility. For example, a mean free path >100 μm is readily observed in the Weyl-II semi-metal $WP_2$[31], which in turn implies that the even the sub-mm sized single crystals used in traditional conductivity measurements are in a quasi-ballistic transport regime. Given the common deviation from circular Fermi surfaces in this materials class, the effects uncovered here are likely to be of relevance to that field. In summary, we believe that much remains to be investigated concerning the physics and potential applicability of directional ballistics.

Acknowledgments
This project was supported by the Max Planck Society and the European Research Council (ERC) under the European Union's Horizon 2020 research and innovation program (MiTopMat - grant agreement No. 715730). M.D.B. acknowledges EPSRC for PhD studentship support through grant number EP/L015110/1. T.S. acknowledges support from the Emergent Phenomena in Quantum Systems initiative of the Gordon and Betty Moore Foundation. Research in Dresden benefits from the environment of the Excellence Cluster ct.qmat. A.S. acknowledges support from an ARCS Foundation Fellowship, a Ford Foundation Predoctoral Fellowship, and a National Science Foundation Graduate Research Fellowship A.S. would like to thank Zack Gomez and Edwin Huang for helpful discussions and Tom Devereaux for letting us use his group cluster. Computational work was performed on the Sherlock cluster at Stanford University and on resources of the National Energy Research Scientific Computing Center, supported by DOE under contract DE_AC02-05CH11231. TS acknowledges support from the Emergent Phenomena in Quantum Systems initiative of the Gordon and Betty Moore Foundation, and from the Natural Sciences and Engineering Research Council of Canada (NSERC), in particular the Discovery Grant [RGPIN-2020-05842], the Accelerator Supplement [RGPAS-2020-00060], and the Discovery Launch Supplement [DGECR-2020-00222].

Data availability
All data needed to evaluate the conclusions in the paper are present in the paper. All raw data underpinning this publication can be accessed in comprehensible ASCII format at DOI: TBD. Source code for the simulations can be found at
https://github.com/dgglab/ballistic_montecarlo.

Author contributions

Competing interests
The authors declare no competing interest.

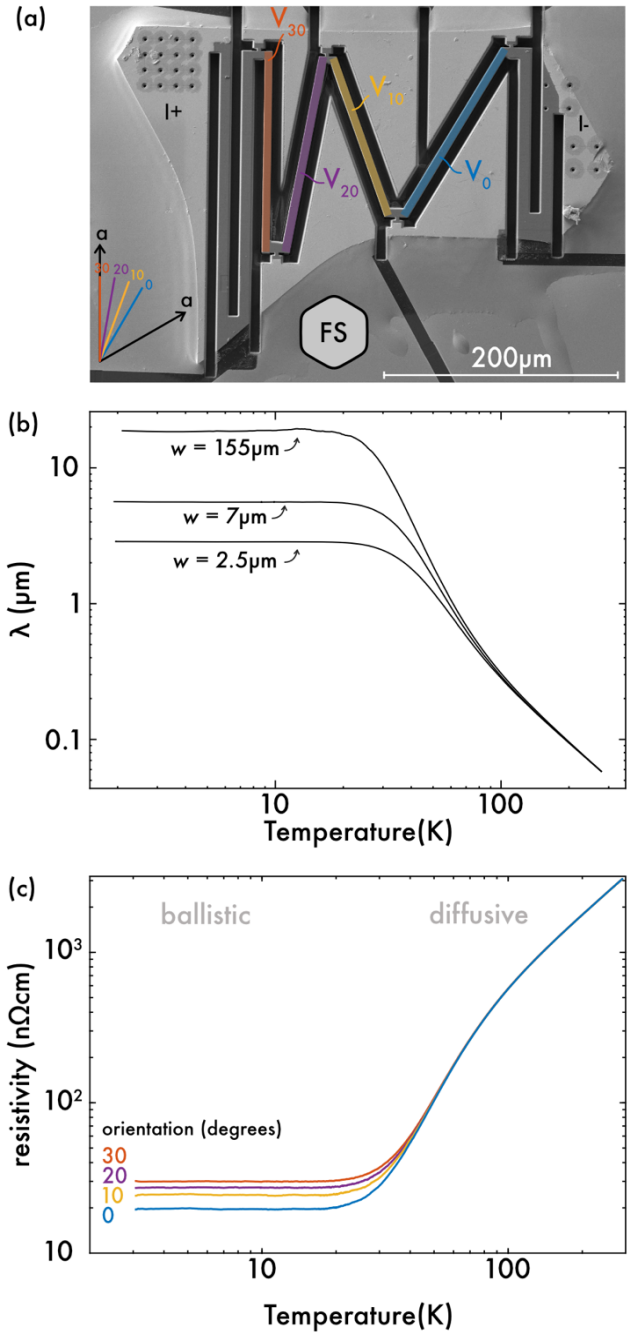

**Fig. 1 | Temperature dependent in-plane transport of PdCoO₂.** (a) One of the PdCoO$_2$ single crystal devices used in this work. A crystal platelet, about 350 μm long and 8 μm thick, has been anchored to a sapphire substrate using two-component Araldite epoxy. A thin layer of titanium/gold (~10nm/150nm) has then been sputtered on top of the device to create electrical contacts to the crystal. In a final step, first the titanium/gold layer is locally removed by FIB etching and subsequently a transport bar is shaped into the crystal with suitable voltage contacts lengthways. To ensure a homogenous current flow throughout the full thickness of the crystal, despite the large resistivity anisotropy of over $^{\rho_c}/_{\rho_a}$ > 2000 below 20K, a long current injection meander has been carved into the device layout[26]. This ensures that the resistivity can be accurately determined along the 4 subsequent transport bars (labelled $V_0$, $V_{10}$, $V_{20}$, $V_{30}$) simultaneously. To this end, five voltage contacts are distributed along the turning-points of the zigzag current path. Additionally, the real-space orientation of the Fermi surface with respect to the crystal is shown. The apparent resistivity

of each bar is defined as the measured voltage divided by the sourced current multiplied by the appropriate geometrical factor. When the sample is in the diffusive regime this corresponds to the bulk resistivity of $PdCoO_2$ but when the ballistic regime is entered it becomes a device-specific quantity. The transport bars of the depicted device have a uniform thickness of 7.8µm, and a width *w* and length *l* of: $w_0$ = 7.3µm, $l_0$ = 170.6µm for $V_0$, $w_{10}$ = 6.5µm, $l_{10}$ = 146.5µm for $V_{10}$, $w_{20}$ = 7.4µm, $l_{20}$ = 172.5µm for $V_{20}$, and $w_{30}$ = 6.7µm, $l_{30}$ = 174µm for $V_{30}$ respectively. (b) The temperature dependent mean free path λ of a $PdCoO_2$ bulk sample (155µm wide; data replotted from reference[14]) and those from the 0° bar pictured in panel (a) at its initial width of 7µm and after narrowing to a width of 2.5µm, all calculated using the standard 2D expression $\rho^{-1} = \frac{e^2}{hd} k_F \lambda$, where *d*=17.73/3Å is the palladium layer separation and $k_F$ = 0.96/Å is the average Fermi momentum. (c) The temperature dependent resistivity of the four transport bars shown in panel (a). In the diffusive regime, all four curves collapse onto the same value, whereas the transport is governed by ballistic effects at lower temperatures. As a result, the residual resistivity is enhanced compared to the bulk value (8 nΩcm[14]) due to boundary scattering. Strikingly, this resistivity enhancement is strongly angle-dependent.

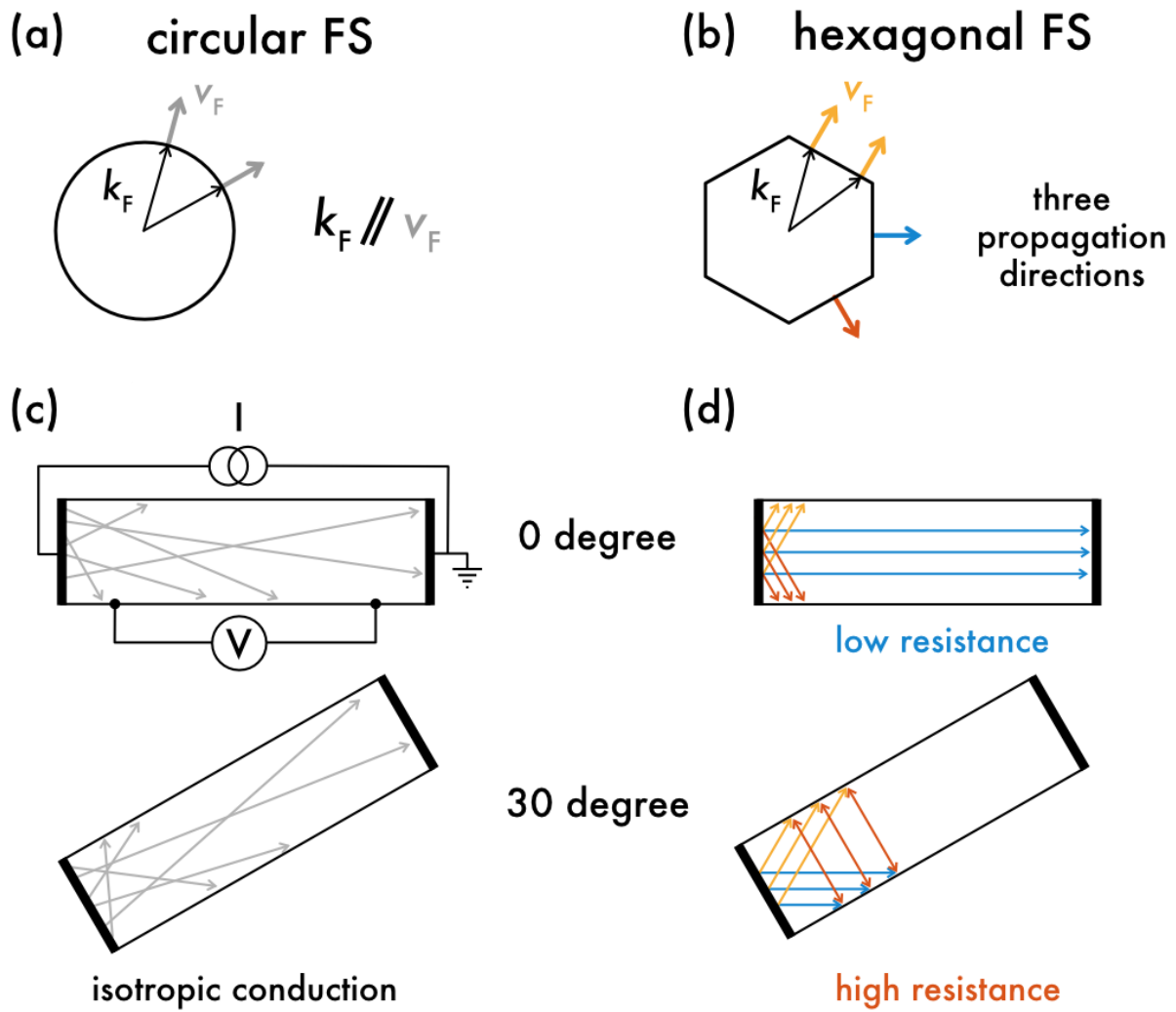

**Fig. 2| Ballistic electron propagation in the case of a circular and a hexagonal Fermi surface.** (a) Whilst the direction of the Fermi velocity $v_F$ is always parallel to the Fermi momentum $k_F$ for a circular Fermi surface, the situation is drastically different in the case of a hexagonal Fermi surface (b). Due to the flat sides of the polygon, there are only three possible directions for the Fermi velocities. This restriction of the electron propagation direction results in a highly anisotropic, directional ballistic transport. The electronic conduction in a 4-point transport bar, fabricated from a material with an isotropic Fermi surface will not depend on the orientation of the bar (c, center). In contrast, for a hexagonal Fermi surface, a bar cut parallel to an electron propagation direction will show a lower resistance than a bar aligned perpendicular to an electron propagation direction.

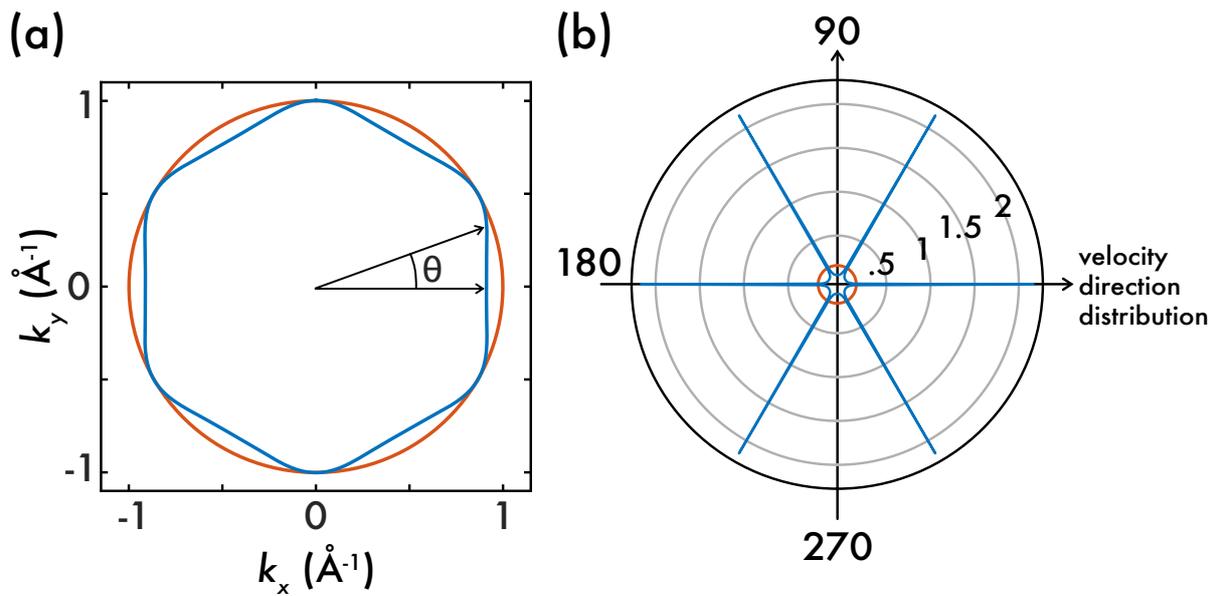

**Fig. 3| The Fermi velocity density and directional ballistics** (a) The Fermi surface of PdCoO$_2$ (blue) compared to a circle (red); at first sight the difference between the two is small. (b) In contrast, the Fermi velocity direction distributions of the two Fermi surfaces are vastly different, with this difference leading to the strong directional ballistic effects that we have measured. While an isotropic Fermi surface (red) leads to a corresponding isotropic Fermi velocity direction distribution, the faceted Fermi surface of PdCoO$_2$ (blue) results in a Fermi velocity direction distribution which is sharply peaked along 6 directions. Note that the velocity direction distribution has been normalized the same for both Fermi surfaces; the sharp spikes in panel b emphasise that each facet of the PdCoO$_2$ Fermi surface contains a large number of states with the same direction of Fermi velocity.

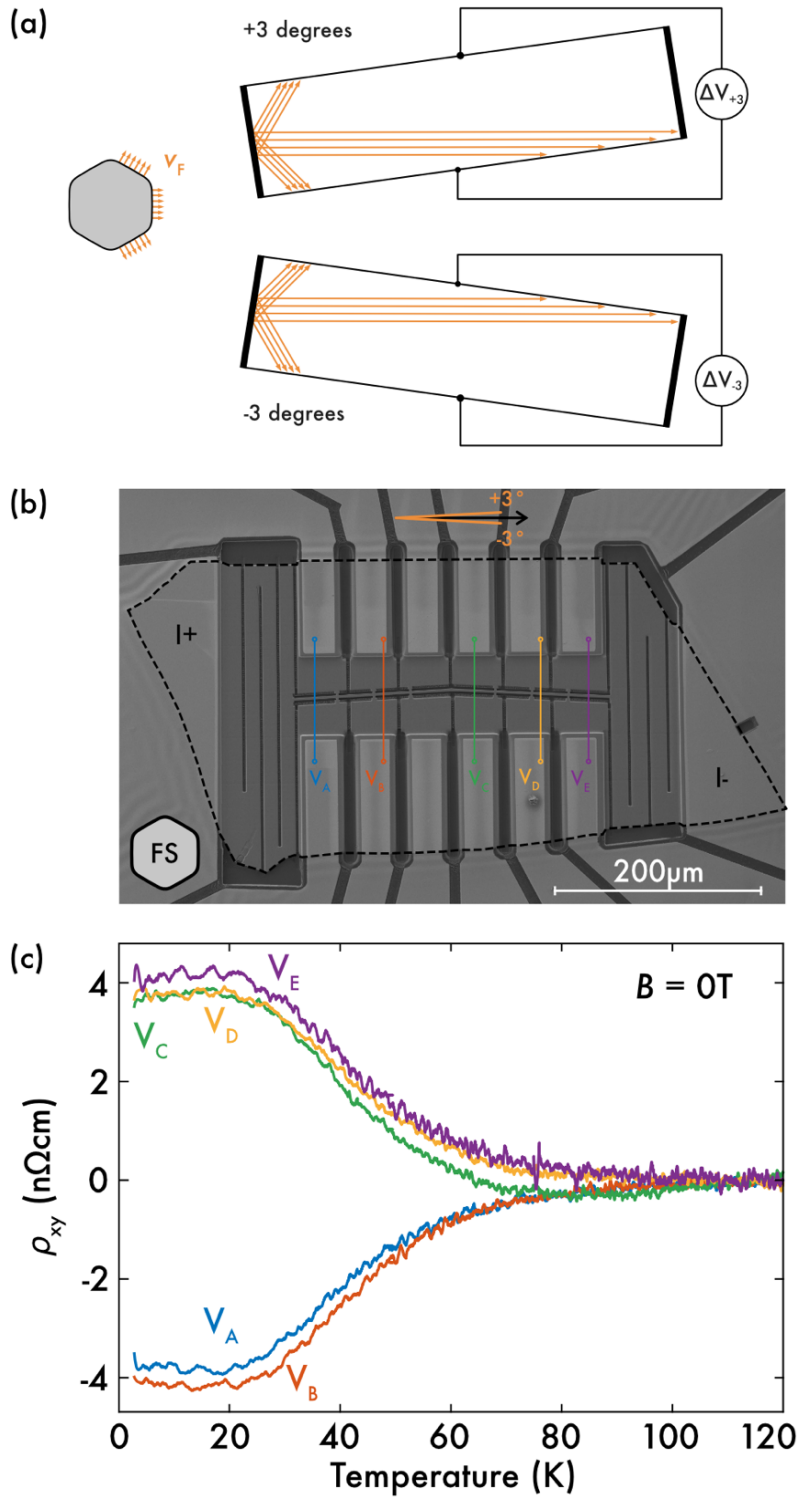

**Fig. 4 | Temperature dependent zero field transverse voltage in PdCoO₂.** (a) Sketch of several representative electronic trajectories in two transport bars cut along low symmetry directions of the crystal, e.g. ±3°. The corresponding orientation of the Fermi surface and Fermi velocities respectively are shown on the left-hand side. (b) A FIB microstructured single crystal of PdCoO₂ with a current path cut along the ±3° directions with 3 pairs of transverse voltage contacts along each orientation. (c) The extracted transverse resistivities increase from the noise floor in the diffusive regime to a finite and anti-symmetric value in the ballistic regime. A longitudinal background stemming from non-ideal

transverse contact alignment (~100nm displacement) has been subtracted. The raw data, sample dimensions and corresponding analysis are presented in Supplementary Note 3.

# Supplementary Information for
"Bachmann *et al.*, Directional ballistic transport in PdCoO$_2$"

## Supplementary Note 1| 2D conductivity tensors on square, triangular and hexagonal crystal lattices

On the Cartesian basis, where $\hat{x}$ and $\hat{y}$ define the plane of the 2D lattice, the generic conductivity tensor $C$ can be written as

$$C = \begin{pmatrix} \sigma_{xx} & \sigma_{xy} \\ \sigma_{yx} & \sigma_{yy} \end{pmatrix}.$$

When the symmetries of the respective crystal system are imposed onto $C$, the number of independent components is reduced. We are concerned with the dihedral point groups D$_n$ with n = 4,6, which have an *n*-fold rotation axis and *n* two-fold reflection axes perpendicular to the rotation axis. The symmetry operations describing the rotation and reflection can be represented in matrix form. The rotation matrix $R(\theta)$, which expresses a counterclockwise rotation through an angle $\theta$ around the out-of-plane axis is given by

$$R(\theta) = \begin{pmatrix} \cos\theta & -\sin\theta \\ \sin\theta & \cos\theta \end{pmatrix}$$

The rotation matrix is orthogonal, i.e. it has the property $R^T R = R R^T = \mathbb{I}$. The rotated conductivity matrix $C'$ is obtained by calculating $C' = R(\theta) \cdot C \cdot R^T(\theta)$.

The matrix describing reflection about the *y*-axis, for instance, is given by

$$S = \begin{pmatrix} -1 & 0 \\ 0 & 1 \end{pmatrix}$$

and the reflected conductivity matrix can be found by calculating $C' = S \cdot C \cdot S^T$. Imposing this reflection symmetry onto the general conductivity tensor $C$ by requiring $C' = C$, we find $\sigma_{xy} = \sigma_{yx} = 0$ and can hence exclude off-diagonal elements.

Further, considering the case of the square lattice, which is invariant under rotation about $\theta = \pi/2$, imposing C$_4$ symmetry on to the conductivity matrix ($C' = R(\pi/2) \cdot C \cdot R^T(\pi/2)$) yields $\sigma_{xx} = \sigma_{yy} \equiv c$ and $\sigma_{xy} = -\sigma_{yx}$. With the reflection symmetry requiring $\sigma_{xy} = \sigma_{yx} = 0$ we find that the general conductivity matrix on a square lattice is given by $C = c \cdot \mathbb{I}$, where $c$ is a scalar and $\mathbb{I}$ is the unitary matrix.

The same formalism can be repeated in the case of a triangular or hexagonal lattice, which are invariant under rotations about $\theta = 2\pi/3$ or $\theta = \pi/3$ respectively. Following the above procedure, one again finds that the conductivity matrices can be expressed by $C = c \cdot \mathbb{I}$.

It is then straightforward to see that in any system, in which the conductivity matrix can be expressed as a scalar multiplied with the identity matrix must have an isotropic conductivity[32].

**Supplementary Note 2| Comparison of the temperature dependent resistivity for devices of different widths.**

The temperature dependent resistivity of $PdCoO_2$ can directly be converted into an estimate for the electron mean free path $\lambda$ using the standard the standard 2D expression[33]:

$$\rho^{-1} = \frac{e^2}{hd} k_F \lambda$$

where $\rho$ is the in-plane resistivity, $e$ is the electron charge, $h$ is the Planck constant, $d=17.73/3$Å is the palladium layer separation, $k_F = 0.96$/Å is the average Fermi momentum.

Here we display and compare the resistivities and resultant mean free path estimations for 3 $PdCoO_2$ devices with widths w = 155μm, 7μm, and 2.5μm. The data for the 155μm wide sample have been replotted from reference[14] and are available at https://edmond.mpdl.mpg.de/imeji/collection/MWrmtAADcp5wH6Yx.

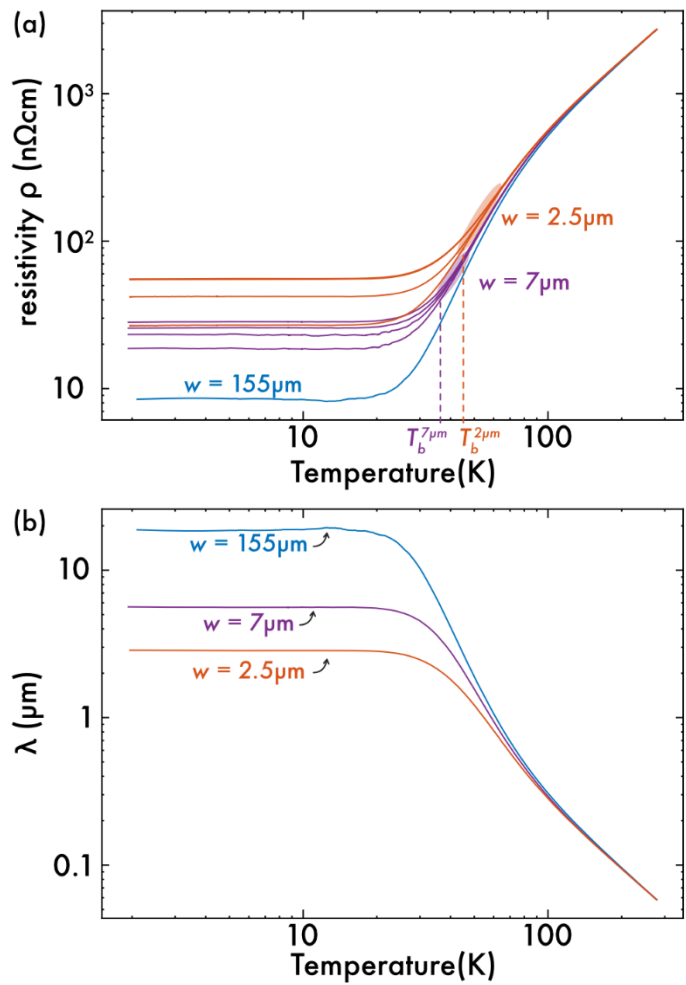

**Fig. S1| Comparison of the temperature dependence of the resistivity and mean free path in $PdCoO_2$ devices of different widths.** (a) The resistivity as a function of temperature for 3 different devices with widths of 2.5μm, 7μm and 155μm (data

for 155µm wide sample taken from reference[14]). For the narrow devices, the resistivities along the 0°, 10°, 20°, and 30° direction are displayed. While the 155 µm device is assumed to be in the bulk limit over the entire temperature range, the narrow devices enter the ballistic regime at low temperatures. The temperatures at which the mean free path $\lambda$ is equal to the device width w is estimated to be 46 K and 34 K for the 2.5 µm and 7 µm wide samples respectively. These temperatures are consistent with an in-plane anisotropy developing in the devices as evidenced by the curves fanning apart at low temperatures. (b) The temperature dependent mean free path $\lambda$ calculated directly from the in-plane resistivity along the 0° direction. While the electrons in the bulk sample (w = 155µm) are not influenced by boundary scattering events, the curves associated with the 2.5µm and 7µm devices are truncated by the device widths rather than the intrinsic bulk mean free path.

**Supplementary Note 3| Raw data and device dimensions of the measurements presented in Fig. .**

Typically, transverse voltage measurements are performed in a magnetic field, where the expected odd voltage response is found by anti-symmetrizing the measured voltage and thereby eliminating any undesired longitudinal (symmetric) component to the signal arising from contact misalignment. Here the measurement is performed in zero field and so this scheme cannot be used. Instead, we first ensured that opposing contacts are aligned as well as possible by in-situ determining the transverse voltage in the ohmic regime at room temperature in the FIB machine and minimizing the signal by using the ion beam to polish the voltage contacts accordingly. From the room temperature values in panels (c) and (d) as well as the device dimensions given listed in panel (b) we can deduce the that contact misalignment is at most 170nm. ($\Delta l = \frac{R}{\rho} \cdot w \cdot h = \frac{1.3\text{m}\Omega}{2.6\mu\Omega\text{cm}} \cdot 2.2\mu m \cdot 1.54\mu m = 170\text{nm}$). Further, in addition to the transverse voltages shown in panel (c), we have also simultaneously determined the longitudinal voltages presented in panel (d). This subsequently allowed us to scale the longitudinal voltages onto the transverse voltages, as indicated by the black curves in panel (e). Finally, the difference between the measured transverse voltages and the scaled longitudinal voltages lead to the sought for pure transverse resistivity presented in panel (f).

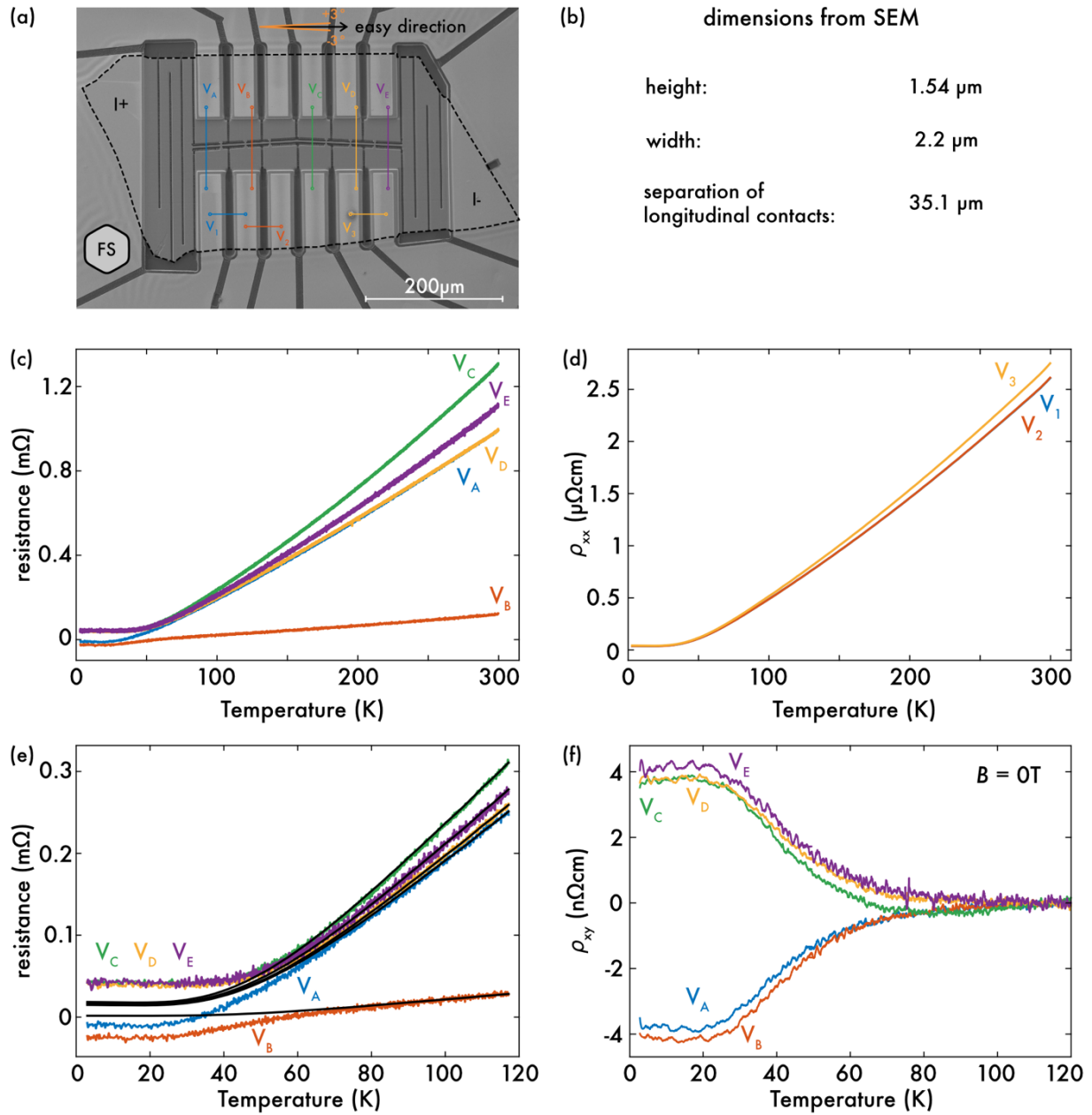

**Fig. S2| Raw data of the measurement presented in figure Fig. .** (a) SEM image of the FIB-fabricated single crystalline $PdCoO_2$ device. (b) Extracted device dimensions from the SEM images. The error bar on all dimensions is ±150nm. (c) Signal recorded between opposing voltage contacts along the current path. No data could be recorded from the pair between $V_B$ and $V_C$ as a contact broke during the initial cool down of the sample. An excitation current of 500µA was used. (d) Measured longitudinal resistivity between the contacts indicated in panel (a). (e) The colored curves are the same as in panel (c). The black curves are the longitudinal voltage signal V1 scaled by a constant factor to fit the transverse curves above 120K. (f) Transverse resistivity after a longitudinal background has been subtracted.

**Supplementary Note 4| Monte Carlo simulations based on Landauer-Büttiker formalism**

We model the electron trajectories according to their semiclassical equations of motion for an out-of-plane magnetic field $\boldsymbol{B} = B\hat{z}$:

$$\hbar v = \frac{\partial \varepsilon}{\partial k}, \qquad \hbar \dot{\boldsymbol{k}} = -e\boldsymbol{E} + eB\hat{z} \times \boldsymbol{v}$$

where $\hbar$ is the reduced Planck's constant, $e$ is the charge of an electron, $v$ is the Fermi velocity, and $\boldsymbol{E}$ is the electric field experienced by the electron. In the ballistic regime, there is negligible electric field in the bulk, therefore we assume that $\boldsymbol{E} = 0$. We take a tight binding approximation of the Fermi surface[15] based on ARPES data[16],

$$\boldsymbol{k}_F(\theta) = k_0 + k_6 \cos(6\theta) + k_{12} \cos(12\theta)$$

where $k_0 = 0.95 \text{Å}^{-1}$, $k_6 = 0.05 \text{ Å}^{-1}$, and $k_{12} = 0.006 \text{ Å}^{-1}$. Because we are not concerned with transit times of the electrons, we can ignore the Fermi velocity $v$.

To model the transport anisotropy, the carriers are injected into a two-dimensional bar of width $w$ and length $L$. Ohmic contacts are created at two ends of the bar which will serve as an injector and a ground. We will assume that the ground is perfect in that any electron colliding with this contact is absorbed and removed from the device. These carriers then follow their semi-classical path[34], ignoring bulk scattering, until interacting with either an edge or ohmic contact of the device. When injecting an electron into the system or scattering from an edge of the device, the probability of injecting into a state $n$ of the discretized Fermi surface is

$$p(n) = cos(\theta(n) - \phi)$$

where $\theta(n) = tan(v_y/v_x)$ is the direction of propagation of the state $n$ and $\phi$ is the angle of the normal to the edge. The Fermi surface is numerically discretized into states separated by constant arclength to remove the probability distribution's dependence on Fermi velocity[34]. The nearly perfectly hexagonal Fermi surface of $PdCoO_2$ has approximately flat edges which cause a high density of states to be injected at fixed angles. In the case of a non-ohmic edge, a carrier is scattered into a new state chosen according to the probability distribution for that edge. To ensure detailed-balance, if the injecting contact absorbs an incident carrier, the carrier is reemitted at a random position along the lead in a randomly chosen allowed state for that edge.

Electrons are injected uniformly across the edge of the contact into states chosen according to the probability distribution for the contact. Therefore, to estimate the two-terminal resistance of the bar, we calculate the forward-propagating flux through a line perpendicular to the channel several times the bars width from the injecting contact. The forward propagating flux is defined as

$$\Phi_{forward} = \frac{1}{w}\left(N_{inject} + \frac{N_{cross} - N_{inject}}{2}\right)$$

where $N_{cross}$ is the number of times charge carriers cross the line and $N_{inject}$ is the number of injected carriers.

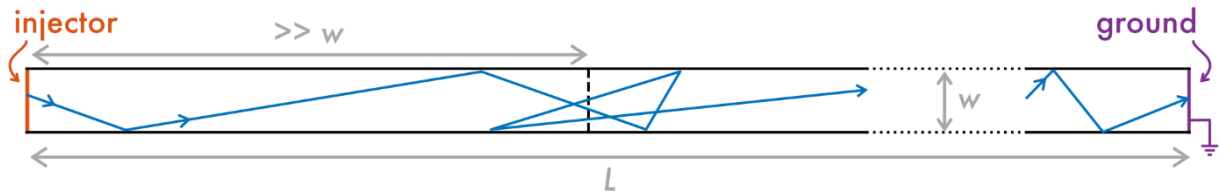

**Fig. S3| Sketch of the simulation setup.** Electrons propagate along a bar of length L and width w. The injector (left) and ground (right) are assumed to be perfect ohmic contacts. A virtual line placed a distance several times the width of the bar away from the injection point is used to calculate to forward propagating flux.

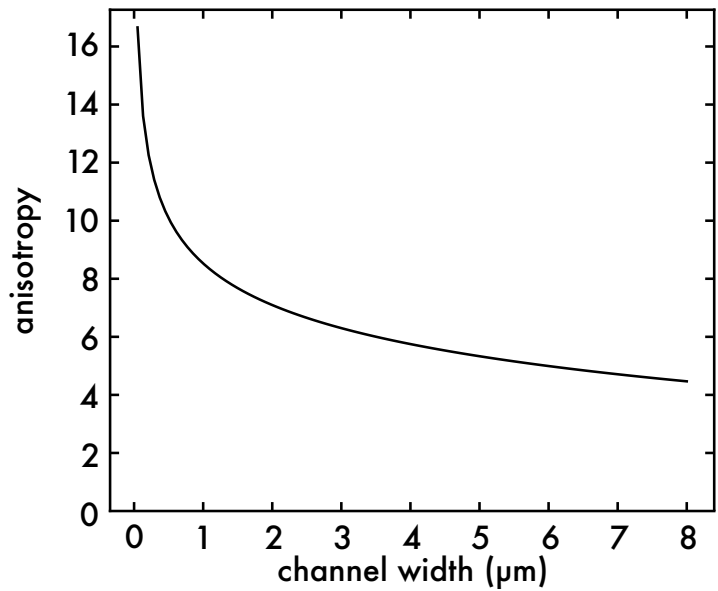

**Fig. S4| Predicted resistivity anisotropy from Monte Carlo Simulations.** Monte Carlo simulations were performed for transport bars along the 0- and 30-degree direction. The ratio of the resistance increases between the 30- and 0-degree direction is shown above and is proportional to the resistive anisotropy. In line with intuition, this anisotropy is predicted to grow rapidly with decreasing channel width.

## Supplementary Note 5| Kinetic calculations for a perfectly hexagonal Fermi surface

We solved the linearized Boltzmann equation in the relaxation time approximation with a perfectly hexagonal Fermi surface. We considered the case of an infinitely long wire along the x direction, and of finite section W along the y direction. The relative direction of the crystalline axis with respect to x is defined by the angle $\theta$, such that $\theta = -\pi/6$ corresponds to the "angle" defined in the main text being 0. Diffuse scattering is assumed at the two boundaries of the wire, i.e. at $y = \pm w/2$.

We denote the out-of-equilibrium distribution of quasi-particles by $\chi(k, y)$, and the mean free path by $\lambda$. It is easy to see that, for the geometry considered, $\chi(k, y)$ takes a uniform value on each of the 6 edges of the Fermi surface. We therefore use the notation $\chi_j(y)$, where j is an index of the Fermi surface edge such that the edge j has velocity $\cos(\phi_j)\hat{x} + \sin(\phi_j)\hat{y}$ with $\phi_j = \left[\frac{\pi}{6}, \frac{3\pi}{6}, \frac{5\pi}{6}, \frac{7\pi}{6}, \frac{9\pi}{6}, \frac{11\pi}{6}\right] - \theta$.

The linearized Boltzmann equation takes the form

$$\sin\phi_j \partial_j \chi_j + \frac{\chi_j}{l} = \frac{\bar{\chi}}{l}$$

with $\bar{\chi} = \frac{1}{6}\sum_j \chi_j$.

The boundary conditions are

$$\chi_{1,2,3}(y = -w/2) = +l\cos\phi_j + \chi_b$$
$$\chi_{1,2,3}(y = +w/2) = +l\cos\phi_j + \chi_t$$

where $\chi_b$ and $\chi_t$ should be chosen so as to satisfy $j_y = 0$.

The density and the currents are given by

$$n = \bar{\chi}$$

$$j_x = \frac{1}{6}\sum_j \chi_j \cos(\phi_j)$$

$$j_y = \frac{1}{6}\sum_j \chi_j \sin(\phi_j)$$

The solution of the Boltzmann equation is

$$\chi_{j=1,2,3} = \chi_j\left(y = -\frac{w}{2}\right)e^{-\frac{y+\frac{w}{2}}{l\sin(\phi_j)}} + \int_{-\frac{w}{2}}^{\infty} dy' \Theta(y-y') e^{-\frac{(y-y')}{l\sin(\phi_j)}} \frac{\bar{\chi}(y')}{l\sin(\phi_j)}$$

$$\chi_{j=4,5,6} = \chi_j\left(y = \frac{w}{2}\right)e^{-\frac{y-\frac{w}{2}}{l\sin(\phi_j)}} - \int_{-\infty}^{W/2} dy' \Theta(y'-y) e^{-\frac{(y-y')}{l\sin(\phi_j)}} \frac{\bar{\chi}(y')}{l\sin(\phi_j)}$$

Proof:

$$\partial_y \chi_{j=1,2,3} = \int_{-\frac{w}{2}}^{\infty} dy' \left( \delta(y-y') e^{-\frac{(y-y')}{l\sin(\phi_j)}} \frac{\bar{\chi}(y')}{l\sin(\phi_j)} + \Theta(y-y') e^{-\frac{(y-y')}{l\sin(\phi_j)}} \frac{\bar{\chi}(y')}{l\sin(\phi_j)} \frac{-1}{l\sin(\phi_j)} \right)$$
$$= \frac{\bar{\chi}(y)}{l\sin(\phi_j)} - \frac{1}{l\sin(\phi_j)} \chi_j$$

For numerical reasons, it is convenient to use

$$F(y_2) = \int_{-\frac{w}{2}}^{y_2} dy' \, e^{-\frac{(y_2-y')}{l\sin(\phi_j)}} \frac{\bar{\chi}(y')}{l\sin(\phi_j)} = e^{(y_1-y_2)/l\sin(\phi_j)} F(y_1) + \int_{y_1}^{y_2} dy' \, e^{-\frac{(y_2-y')}{l\sin(\phi_j)}} \frac{\bar{\chi}(y')}{l\sin(\phi_j)}$$

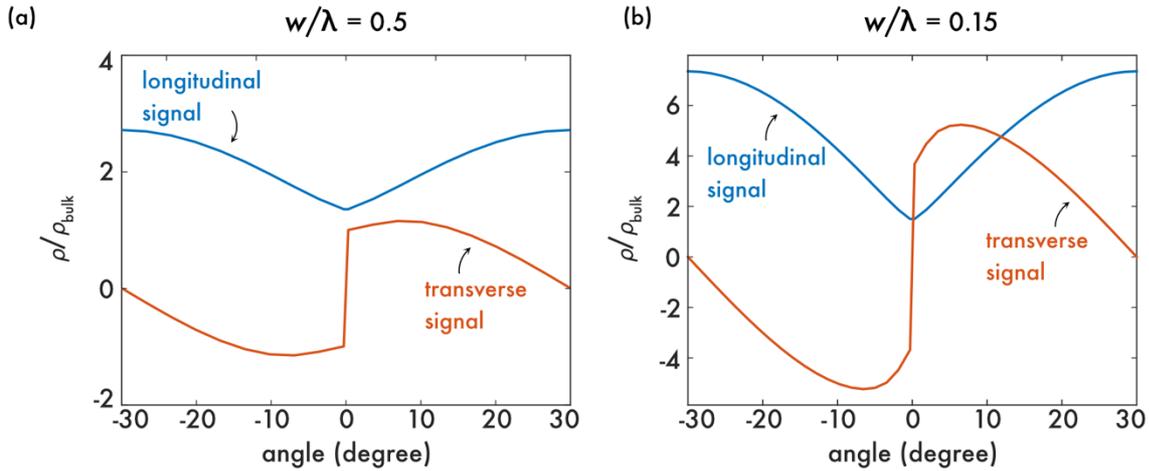

**Fig. S5|Calculated longitudinal and transverse resistivity anisotropy for a perfectly hexagonal Fermi surface.** for two ratios of $w/\lambda$, specifically (a) $w/\lambda = 0.5$ and (b) $w/\lambda = 0.15$. The anisotropies in the longitudinal anisotropy are 1.9 and 5, both overestimates compared to the experimentally measured values, with the discrepancy quite large for the narrower channel. This is likely to be a result of using the idealized perfect hexagon for the Fermi surface instead of a more realistic one with rounded corners. Unsurprisingly this also results in an overestimate of the ratio of transverse to longitudinal resistivity, predicted to lie in the range 0.8-2.6 for channels of the dimension shown in Fig. 3, compared with our measured value of 0.3. Nevertheless, the qualitative trends of our observations are well captured by the calculations.

**Supplementary Note 6| Calculation of the Fermi velocity direction distribution**

Here we describe the derivation of the angular distribution of the Fermi velocity direction based on an angularly parametrized Fermi surface.

From precise ARPES measurements[35], the in-plane Fermi surface of $PdCoO_2$ is well known and can be expressed as a periodic function of angle $\varphi$:

$$\begin{pmatrix} k_F^x \\ k_F^y \end{pmatrix} = k_0 \begin{pmatrix} \cos\varphi \\ \sin\varphi \end{pmatrix} + k_6 \cos 6\varphi \begin{pmatrix} \cos\varphi \\ \sin\varphi \end{pmatrix} + k_{12} \cos 12\varphi \begin{pmatrix} \cos\varphi \\ \sin\varphi \end{pmatrix}$$

where the values $k_0$ = 0.9518Å$^{-1}$, $k_6$ = 0.0444Å$^{-1}$, $k_{12}$ = 0.0048Å$^{-1}$ are taken from[35]. The corresponding curve is displayed in figure S6.

Next, the Fermi surface must be expressed by points separated by constant arc length. The arc length, s, as a function of polar angle is given by $s(\varphi) = \sqrt{(\partial_x k_F^x(\varphi))^2 + (\partial_y k_F^y(\varphi))^2}$. With this, the Fermi surface curve can be interpolated, such that it is spanned by points of equal arc length. This is straightforward to implement numerically. Next, we set out to find the distribution of the Fermi velocity direction, θ, in Fig. S6, as a function of polar angle φ.

In general, the direction of a normal vector at any point along a parametric curve **r**($\varphi$) = (**x**($\varphi$), **y**($\varphi$)) can be found by first taking the derivative ∂**r**/∂$\varphi$ = (∂**x**/∂$\varphi$, ∂**y**/∂$\varphi$), which yields the tangent vector to the curve and then rotating it clockwise by 90 degrees. This results in a vector $\widetilde{v_F}$ = (∂**y**/∂$\varphi$, -∂**x**/∂$\varphi$) which is parallel to the Fermi velocity. Finally, the velocity direction is simply given by the angle $\theta$ of the rotated tangent vector, which can be found by calculating $\theta(\varphi) = \tan^{-1}\left(-\frac{\partial \mathbf{x}}{\partial \varphi} / \frac{\partial \mathbf{y}}{\partial \varphi}\right)$. This calculation is displayed graphically in Fig. S6.

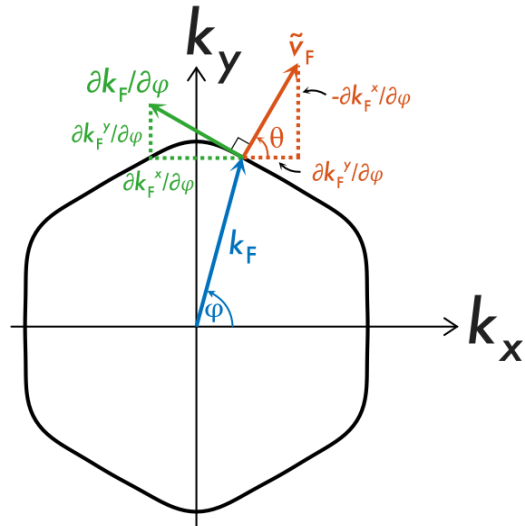

**Fig. S6|Derivation of the Fermi velocity direction distribution.** In order to determine the Fermi velocity distribution the direction of the Fermi velocity, $\theta$, as a function of polar angle $\varphi$ must be calculated.

Finally, the angular Fermi velocity direction distribution is found by evaluating the function $\theta(\varphi)$ at equidistant arc lengths along the Fermi surface and displaying the normalized probability density as a function of polar angle, as is presented in Fig. 3.